\documentclass{emulateapj}
\usepackage{natbib}
\newcommand{\ergseccm}{erg\,sec$^{-1}$\,cm$^{-2}$}
\newcommand{\etal}{et\,al.}
\newcommand{\halpha}{H$\alpha$}
\newcommand{\gsim}{\raise0.3ex\hbox{$>$}\kern-0.75em{\lower0.65ex\hbox{$\sim$}}}
\newcommand{\fnu}{{\it f}$_{\rm \nu}$}
\newcommand{\kms}{km\,s$^{-1}$}
\newcommand{\lsim}{\raise0.3ex\hbox{$<$}\kern-0.75em{\lower0.65ex\hbox{$\sim$}}}
\newcommand{\msun}{M$_{\odot}$}
\newcommand{\mm}{$\mu$m}
\newcommand{\pom}{\,$\pm$\,}
\newcommand{\sings}{{\it SINGS}}
\newcommand{\spitzer}{{\it Spitzer}}
\newcommand{\HI}{H~{\sc i}}
\newcommand{\HII}{H~{\sc ii}}
\newcommand{\snu}{S$_{\nu}$} 
\begin{document}     
\slugcomment{ApJ Letters, in press; Submitted 23 June 2005; Accepted 28 July
2005}
\shorttitle{{\it Spitzer} Observations of IC\,2574}
%-----------------------------------------------------------------------------%
\title{Spitzer Observations Of The Supergiant Shell Region In IC 2574}
%-----------------------------------------------------------------------------%

\author{John M. Cannon\footnotemark[1],
Fabian Walter\footnotemark[1],
George J. Bendo\footnotemark[2],
Daniela Calzetti\footnotemark[3],
Daniel A. Dale\footnotemark[4],
Bruce T. Draine\footnotemark[5],
Charles W. Engelbracht\footnotemark[2],
Karl D. Gordon\footnotemark[2],
George Helou\footnotemark[6],
Robert C. Kennicutt, Jr.\footnotemark[2],
Eric J. Murphy\footnotemark[7],
Michele D. Thornley\footnotemark[8],
Lee Armus\footnotemark[6],                                       
David J. Hollenbach\footnotemark[9],
Claus Leitherer\footnotemark[3],
Michael W. Regan\footnotemark[3],
H{\' e}l{\` e}ne Roussel\footnotemark[6],
Kartik Sheth\footnotemark[6]}

\footnotetext[1]{Max-Planck-Institut f{\"u}r Astronomie, K{\"o}nigstuhl 17, 
D-69117, Heidelberg, Germany; cannon@mpia.de, walter@mpia.de}
\footnotetext[2]{Steward Observatory, University of Arizona, 933 North Cherry 
Avenue, Tucson, AZ 85721}
\footnotetext[3]{Space Telescope Science Institute, 3700 San Martin Drive, 
Baltimore, MD 21218}
\footnotetext[4]{Department of Physics and Astronomy, University of Wyoming, 
Laramie, WY 82071}
\footnotetext[5]{Princeton University Observatory, Peyton Hall, Princeton, NJ 
08544}
\footnotetext[6]{California Institute of Technology, MC 314-6, Pasadena, CA
91101}
\footnotetext[7]{Department of Astronomy, Yale University, New Haven, CT 
06520}
\footnotetext[8]{Department of Physics, Bucknell University, Lewisburg, PA 
17837}
\footnotetext[9]{NASA/Ames Research Center, MS 245-6, Moffett Field, CA, 
94035}

\begin{abstract}
%-----------------------------------------------------------------------------%

We present spatially resolved {\it Spitzer} imaging of the supergiant shell 
region of the M81 group dwarf galaxy IC\,2574 obtained as part of the {\it 
Spitzer Infrared Nearby Galaxies Survey}. This region harbors one of the best 
nearby examples of a kinematically distinct \HI\ shell, with an associated 
remnant stellar cluster; the shell is initiating sequential star formation as 
it interacts with the surrounding interstellar medium. This region dominates 
the infrared luminosity of IC\,2574 and is spatially resolved in all {\it 
Spitzer} imaging bands.  We study the differences in dust temperature as a 
function of local environment and compare local star formation rates as 
inferred from \halpha\ and total infrared luminosities.  We find that the 
strong \halpha\ sources are associated with regions of warm dust; however, the 
most luminous infrared and \halpha\ sources are not necessarily
co-spatial. The coolest dust is found in the regions farthest from the rim of 
the shell; these regions show the best agreement between star formation rates 
derived from \halpha\ and from total infrared luminosities (although 
discrepancies at the factor of 3--4 level still exist).  There is considerable
variation in the radio-far infrared correlation in different regions 
surrounding the shell.  The low dust content of the region may influence the 
scatter seen in these relations; these data demonstrate that the expanding 
shell is dramatically affecting its surroundings by triggering star formation 
and altering the dust temperature.

\end{abstract}						


%-----------------------------------------------------------------------------%
\section{Introduction}
\label{S1}
%-----------------------------------------------------------------------------%

One of the most dramatic effects of vigorous star formation (SF) is the 
creation 
of holes and shells in the interstellar medium (ISM).  It is commonly proposed 
that these structures are caused by feedback from massive stars (stellar winds 
and Type II SNe; e.g., {Tenorio-Tagle \& Bodenheimer 
1988}\nocite{tenoriotagle88}), though alternative scenarios do exist (high 
velocity cloud impacts, disk instabilities, turbulence, ram pressure
stripping; e.g., {S{\' a}nchez-Salcedo 2002}\nocite{sanchezsalcedo02}). 
Starburst regions and sites of massive cluster formation provide spatially and 
temporally concentrated feedback that can create the largest of these 
structures.  Holes and shells are therefore a direct, observable signature of 
the deposit of energy from stars into the ISM. 

These structures are found in a great variety of environments, from the Milky
Way disk \citep[e.g.,][]{mccluregriffiths02} to the ISM of dwarf galaxies
\citep[e.g.,][]{walter99a}. After being formed by the combined effects of 
SNe and stellar winds, the evolution of holes and shells will 
have a strong environmental dependence; in normal spiral disks, these 
structures are erased by turbulent motions and rotational shear on time scales 
of $\sim$ 10$^7$ yr. On the other hand, in dwarf galaxies (typically 
displaying solid-body rotation), rotational shear will not play a major role, 
and holes and shells may remain coherent until pressure equilibrium is 
re-established with the local ISM \citep[e.g.,][]{elmegreen00}. Indeed,
studies of active, nearby dwarf galaxies show that these systems are permeated 
with holes and shells to the \HI\ resolution limit [e.g., the Large Magellanic 
Cloud ({Kim \etal\ 1999}\nocite{kim99}); the Small Magellanic Cloud 
({Stanimirovic \etal\ 1999}\nocite{stanimirovic99}); and IC\,10 ({Wilcots \& 
Miller 1998}\nocite{wilcots98}), to name just a few].

\begin{figure*}[!ht]
\plotone{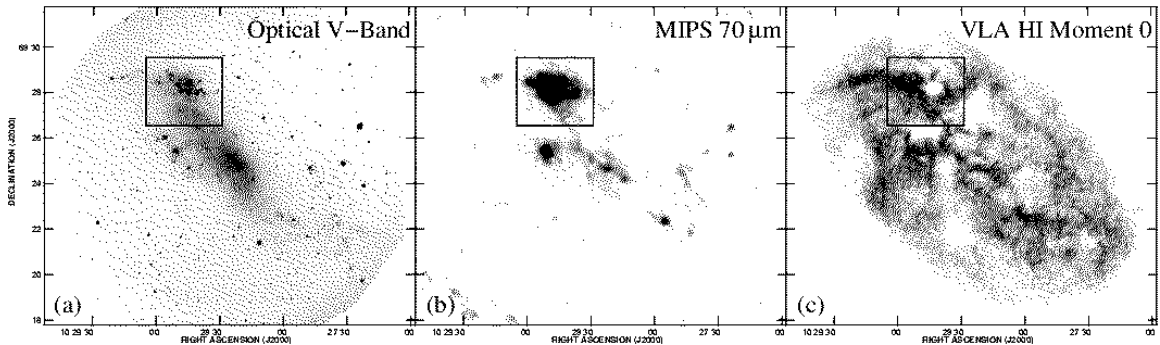}
\caption{Optical V-band (a), MIPS 70 \mm\ (b) and \HI\ (c) images of IC\,2574.
Note that the SGS region dominates the far-IR luminosity of IC\,2574, though 
emission is associated with other star formation regions. The box in each 
field denotes the area shown in Figure~\ref{figcap2}, and is $\sim$ 
4.4\,$\times$\,3.8 kpc at the adopted distance.}
\label{figcap1}
\end{figure*}

IC\,2574 is a comparatively large (optical disk $\sim$ 18 kpc diameter),
relatively low-metallicity [12$+$log(O/H) $\simeq$ 8.15, or $\sim$ 30\% 
Z$_{\odot}$; {Miller \& Hodge 1996}\nocite{miller96}] dwarf galaxy in the M81 
group that is undergoing active current SF (\halpha-derived SFR 
$\simeq$ 0.09 \msun\,yr$^{-1}$; {Miller \& Hodge 1994}\nocite{miller94}, 
{Kennicutt 1998}\nocite{kennicutt98}).  The galaxy hosts a multitude of \HI\ 
holes and shells \citep{walter99b}, which at the distance of 4.0 Mpc 
\citep{karachentsev02} provide a unique opportunity to study
the process of SF and to observe its effects on the surrounding 
ISM.  The most dramatic ``supergiant shell'' 
\citep[SGS; ][]{walter98,walter99b} is expanding at $\sim$ 25 \kms, has a 
diameter of $\sim$ 790 pc, a kinematic age of 15.8 Myr, an \HI\ mass of $\sim$ 
8.2$\times$10$^{5}$ \msun, and requires an energy input of $\sim$ 
5.8$\times$10$^{52}$ erg. As shown in \citet{stewart00}, the progenitor
stellar cluster is interior to the shell, and the expanding structure is 
igniting sequential SF on the shell rim. 

%-----------------------------------------------------------------------------%
\section{Observations and Data Reduction}
\label{S2}
%-----------------------------------------------------------------------------%

For an overview of the {\it Spitzer Infrared Nearby Galaxies Survey} (\sings) 
observational strategies, see \citet{kennicutt03}.  IC\,2574 was observed for 
71 minutes in IRAC mosaicing mode on 28 and 29 October, 2004; MIPS scan 
mapping mode observations were obtained on 1, 3 November, 2004, for a total of
127.6 minutes.  All data were processed by the \sings\ pipelines.  The IRAC 
pipeline processes basic calibrated data images; flux levels are uncertain at 
the $\sim$ 10\% level. The MIPS Instrument Team Data Analysis Tool 
\citep{gordon05} was used to process the MIPS data.  Systematic uncertainties 
(e.g., detector nonlinearities, time-dependent responsivity variations, 
background removal, etc.) limit the absolute flux calibration to $\sim$ 10\% 
at 24 \mm\ and to $\sim$ 20\% at 70 and 160 \mm. Fluxes were measured after 
convolution to the (38\arcsec\ FWHM) 160\mm\ MIPS beam, using kernels derived 
from observations of a bright star (IRAC) or from STinyTim models, smoothed to 
match the observed PSFs (assuming a 25 K black body, suitable for the dust 
temperatures derived; MIPS).  

%-----------------------------------------------------------------------------%
\section{Multiwavelength Emission from the SGS}
\label{S3}
%-----------------------------------------------------------------------------%

The SGS region provides the bulk ($\simeq$ 50\% at 24 and 70 \mm) of the total 
IR (TIR) luminosity of IC\,2574 at wavelengths longer than $\sim$ 5 \mm.
In Figure~\ref{figcap1} we present a 
comparison of the total galaxy emission in the optical V band, the MIPS 70 
\mm\ band, and the \HI\ spectral line.  Note in Figure~\ref{figcap1}(c) that 
IC\,2574 contains numerous \HI\ shells ({Walter \& Brinks 
1999}\nocite{walter99b} identify 48 holes and shells in the ISM; the SGS 
studied here corresponds to \# 35 from that study).

\begin{figure*}[!ht]
\plotone{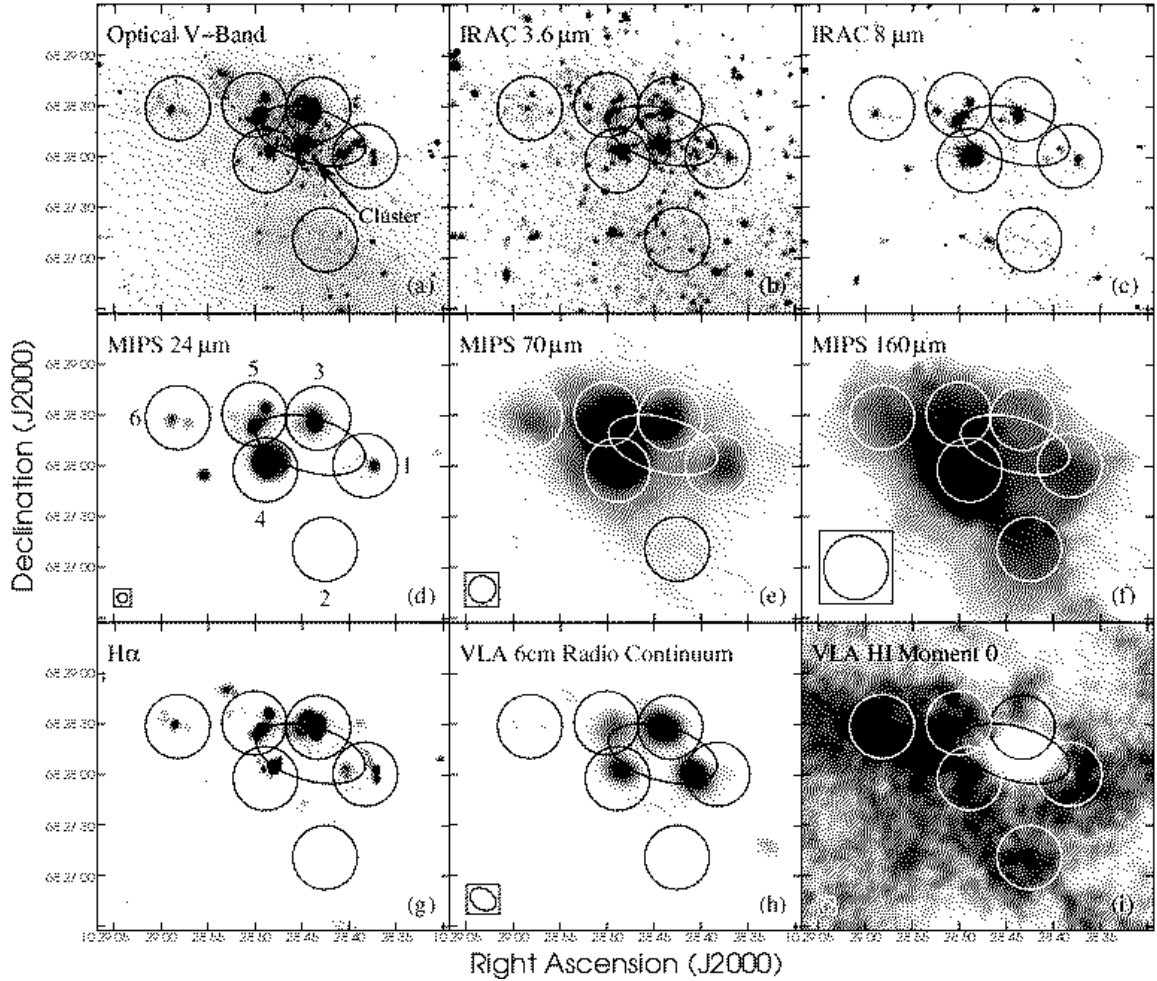}
\caption{Images of the SGS region at nine different wavelengths (see 
Figure~\ref{figcap1} for location): optical V-band (a); IRAC 3.6 and 8 \mm\
(b, c); MIPS 24, 70 160 \mm\ (d, e, f); continuum-subtracted \halpha\ (g); 
6\,cm radio continuum (h); \HI\ moment zero image. The location of circular 
apertures SGS\,1 - 6 are overlaid, and labeled in (d); the SGS is shown as an 
ellipse in each frame. In frames (d), (e), (f), and (i), the beam sizes (FWHM) 
are shown as boxed circles at bottom left.}
\label{figcap2}
\end{figure*}

In Figure~\ref{figcap2} we present images of the SGS region at nine  
different wavelengths.  Note from the optical (V-band; Figure~\ref{figcap2}a) 
and near-IR (IRAC 3.6 \mm; Figure~\ref{figcap2}b) images that the progenitor 
stellar cluster lies directly interior to the \HI\ shell \citep[see 
also][]{stewart00}; this is one of the clearest examples of a kinematically 
distinct gaseous shell with the parent cluster still visible.  By 8 \mm\
(see Figure~\ref{figcap2}c) the spectral energy distribution of the cluster 
has fallen below the detection limit, and emission from hot dust and gas 
dominates; variations in emission in the MIPS bands (Figures~\ref{figcap2}d, 
e, f) indicate a wide variety of dust temperatures and spectral energy 
distributions. Note that the shell morphology is still evident at 70 and 
160 \mm; the diffuse emission in the shell is most likely caused by the 
MIPS PSF profiles, which spread flux from high surface brightness regions 
onto arcminute scales (i.e., a few times the MIPS 160 \mm\ PSF FWHM).
Comparison of the MIPS, \halpha\ and radio continuum images 
[Figures~\ref{figcap2}(g, h)] shows a wide variation in the relative ratios. 
Finally, the \HI\ distribution [Figure~\ref{figcap2}(i)] shows the \HI\ shell
very clearly; it is expanding into a non-uniform medium that may partially 
explain the variety of dust properties around the shell rim.  

Flux densities were extracted in the apertures shown in Figure~\ref{figcap2} 
(see Table~\ref{t1}); these regions were selected to encompass the mid-IR, 
far-IR, \halpha, and \HI\ emission peaks with the minimum number of apertures 
and amount of overlap (note that some apertures contain distinct emission 
properties at different wavelengths; e.g., SGS\,1 shows a pronounced
difference in \halpha\ and radio continuum morphologies). The size of these 
apertures corresponds to the FWHM (38\arcsec\ $=$ 740 pc) of the MIPS 160 \mm\ 
beam. Aperture correction factors have not been applied to the values shown in 
Table~\ref{t1}; the extent of the apertures for all wavelengths up to the MIPS 
70 \mm\ band (sampling more than 2, 6, 16, and 36 times the FWHM at MIPS 70 
\mm, MIPS 24 \mm, IRAC 8 \mm, and IRAC 3.6 \mm, respectively) should imply 
relatively small aperture correction effects at these wavelengths.  There will 
be some aperture effects at MIPS 160 \mm\ and potentially at 70 \mm\ as well, 
though the quantification of these factors depends on many parameters 
(including the distribution of light within the aperture, location and 
brightness of neighboring sources, and many others). 

%-----------------------------------------------------------------------------%
\section{Varying Dust Conditions in the SGS Region}
\label{S4}
%-----------------------------------------------------------------------------%

The characteristics of the dust change dramatically between apertures. There 
are appreciable variations in the dust temperature; 
\fnu(70\,\mm)/\fnu(160\,\mm) varies by a factor of $\sim$ 3, with the warmest
dust in regions SGS\,3 and SGS\,5.  Fitting blackbody functions modified by a 
$\lambda^{-2}$ emissivity to the 70 and 160 $\mu$m MIPS images \citep[see, 
e.g., ][]{bianchi99}, these fluctuations correspond to a temperature range of 
$\sim$ 23-29 K (averaged over each aperture; see Table~\ref{t1}).  Note that 
the warmest dust is found in regions that have the highest \halpha\ flux; the 
coolest dust is found in region SGS\,2, which is at the largest distance from 
the SGS itself, and also shows the lowest \halpha\ flux. Interestingly, 
region SGS\,4, which shows the highest flux density in all \spitzer\ bands, 
does not show the highest \halpha\ flux.  

Comparing the far-IR, \halpha, and \HI\ morphologies, it is evident that all 
regions except SGS\,3 occupy areas with high \HI\ surface brightness (\HI\ 
column densities \gsim\ 2$\times$10$^{21}$ cm$^{-2}$).  SGS\,3 is the region 
with the largest \halpha\ and 6\,cm radio continuum fluxes, arguing for strong 
active SF, and also the highest \fnu(70\,\mm)/\fnu(160\,\mm) dust 
temperature ratio.  Given the age of the shell ($\simeq$ 15 Myr) and the 
expected lifetimes of \HII\ regions and associated thermal radio continuum 
emission (\lsim\ 30 Myr), this shell may be initiating rapid SF 
that quickly disperses the local gas supply.  

\begin{deluxetable*}{lccccccc}[!ht]
\tabletypesize{\scriptsize}
\tablecaption{Dust Emission and Derived Properties in the IC\,2574 Super Giant 
Shell\tablenotemark{a}}
\tablewidth{0pt}
\tablehead{
\colhead{Parameter}         
&\colhead{SGS 1}
&\colhead{SGS 2} 
&\colhead{SGS 3}
&\colhead{SGS 4}
&\colhead{SGS 5}
&\colhead{SGS 6}
&\colhead{SGS Total}}
\startdata
\vspace{0.1 cm}
$\alpha$ (J2000) &10:28:38.275 &10:28:42.559 &10:28:43.301 &10:28:48.912 
&10:28:50.102 &10:28:58.260 &10:28:44.1\\
$\delta$ (J2000) &68:28:00.44 &68:27:10.98 &68:28:28.32 &68:27:57.83 
5&68:28:30.76 &68:28:28.66 &68:28:12.6\\
\halpha\ Flux\tablenotemark{b}
&37\pom6 
&7.5\pom1.3 
&102\pom20 
&49\pom7 
&57\pom9 
&17\pom3 
&380\pom60\\
\halpha\ Convolved Flux\tablenotemark{c}
&28 \pom6 
&8.5\pom1.3 
&66\pom20 
&37 \pom7 
&48 \pom9 
&12 \pom3 
&380\pom60\\
IRAC 3.6\,\mm\ Flux Density
&1.4\pom0.2
&1.6\pom0.2
&1.9\pom0.3
&2.0\pom0.3
&1.8\pom0.3
&1.1\pom0.2
&28\pom4\\
IRAC 4.5\,\mm\ Flux Density 
&1.3\pom0.2
&1.4\pom0.2
&1.7\pom0.3
&1.9\pom0.3
&1.7\pom0.3
&1.2\pom0.2
&29\pom4\\
IRAC 8.0\,\mm\ Flux Density 
&1.1\pom0.2
&1.1\pom0.2
&1.5\pom0.2
&2.9\pom0.4
&1.9\pom0.3
&0.9\pom0.1
&22\pom3\\
MIPS 24\,\mm\ Flux Density  
&5.3\pom0.8
&2.6\pom0.4
&12\pom2
&24\pom4
&15\pom2
&4.6\pom0.7
&100\pom20\\
MIPS 70\,\mm\ Flux Density  
&110\pom20
&70\pom20
&160\pom30
&250\pom50
&240\pom50
&97\pom20
&1900\pom400\\
MIPS 160\,\mm\ Flux Density 
&150\pom40
&160\pom40
&140\pom40
&290\pom70
&250\pom60
&160\pom40
&2600\pom650\\
6\,cm Flux Density
&0.63\pom0.07
&0.046\pom0.01
&0.88\pom0.09
&0.53\pom0.06
&0.50\pom0.06
&0.13\pom0.02
&3.03\pom0.4\\ 
F(\halpha)/F(TIR)
&0.044
&0.011
&0.090
&0.024
&0.033
&0.021
&0.026\\ 
F(\halpha)/F(TIR)\tablenotemark{c}
&0.034
&0.012
&0.058
&0.019
&0.028
&0.014
&0.026\\ 
T$_{\rm DUST}$ (K)
&26\pom3
&23\pom3
&29\pom3
&26\pom3
&28\pom3
&25\pom3
&N/A\\ 
q (TIR/Radio)\tablenotemark{d}
&2.55
&3.61
&2.53
&3.01
&2.96
&3.23
&3.11\\ 
\enddata
\enddata
\label{t1}\vspace{-0.4 cm}
\tablenotetext{a}{All values listed in units of mJy, unless otherwise noted.
\spitzer\ flux densities were derived without aperture corrections, using 
images convolved to the MIPS 160\mm\ resolution, with foreground stars removed 
in the IRAC and MIPS 24\mm\ bands.}\\
\tablenotetext{b}{Derived from the work of \citet{miller94}; units of  
10$^{-14}$\ergseccm.}
\tablenotetext{c}{Using the \halpha\ image convolved to the MIPS 160 \mm\ 
FWHM $=$ 38\arcsec.}
\tablenotetext{d}{The value of q in the radio-far IR relation, derived using
the total IR flux and the radio continuum flux at 6\,cm. These calculations 
assume a mix of thermal and nonthermal emission (i.e., $\alpha = -$0.7, where 
\snu\ $\sim\ \nu^{\alpha}$) for the extrapolation from 20 to 6\,cm flux 
densities.}
\end{deluxetable*}

%-----------------------------------------------------------------------------%
\subsection{Variations in the IR vs. \halpha\ Ratio}
\label{S4.1}
%-----------------------------------------------------------------------------%

In principle, all UV photons within dusty environments will
be absorbed and re-radiated in the IR; thus, there should exist a correlation 
between the IR luminosity and other SFR indicators, such as \halpha\ emission.
However, this simple scenario can become complicated in different 
environments, with dependencies on stellar populations, dust content, etc.
(see {Kennicutt 1998}\nocite{kennicutt98} for a more detailed discussion). In 
the case of (typically low-metallicity) dwarf star-forming galaxies such as 
IC\,2574, the correlation between various SFR indicators may be especially 
complicated, given their low dust contents and IR luminosities.  Dust 
extinction effects (in both an absolute and differential sense) on the 
\halpha\ fluxes in the SGS are minimal, based on two independent lines of 
evidence:  first, the UV imaging study of this region by \citet{stewart00} 
shows typical line-of-sight reddening values in the wavelength range around 
\halpha\ of A$_{\rm R} \simeq$ 0.15 mag, with a maximum of $\sim$ 0.3 mag; 
second, new {\it HST}/ACS imaging and color-magnitude diagram analysis of this
region (E.D. Skillman, private communication) easily separates the blue 
helium-burning and main sequence stars, which at this metallicity are 
separated by $\simeq$ 0.2 magnitudes in (V$-$I) color. 

We find that the L(\halpha)/L(TIR) ratio is systematically higher (by factors 
of $\sim$ 10; see Table~\ref{t1}) in regions that are bright in \halpha\ 
compared to more quiescent regions. If one were to naively convert the 
\halpha\ and IR fluxes to SFRs [using the relations of {Kennicutt 
(1998)}\nocite{kennicutt98} and {Dale \& Helou (2002)}\nocite{dale02}], the 
SFR(\halpha) values would be systematically higher (by factors of $\sim$ 10) 
than those derived from the total IR luminosity in the active regions.  Since 
the \citet{kennicutt98} FIR calibration assumes complete absorption of the 
starlight, this factor-of-ten difference implies a typical H$\alpha$ 
extinction in these regions of about 10\% (A(H$\alpha$) $\sim$ 0.1 mag). This 
is roughly consistent with the range of values cited above, and suggests 
caution in the application of SFR relations based on total IR luminosities in
low-metallicity environments. Note that the variations in the
L(\halpha)/L(TIR) ratio are not due only to varying spatial resolutions
between \halpha\ and 160 \mm; we tested the severity of this effect by 
convolving the \halpha\ image to the 160 \mm\ resolution, and find that the 
L(\halpha)/L(TIR) ratio still varies by factors of $\sim$ 5 within the SGS
region (see Table~\ref{t1}).

%-----------------------------------------------------------------------------%
\subsection{Variations in the Radio-Far IR Correlation}
\label{S4.2}
%-----------------------------------------------------------------------------%

Assuming that the total IR fluxes (derived by applying equation 4 from {Dale 
\& Helou 2002}\nocite{dale02}) and the radio continuum emission are related 
via a constant value [q $\propto$ log(S$_{\rm TIR}$/S$_{\rm RC}$); see, e.g.,
{Bell (2003)}\nocite{bell03a} for details], we derive strong variations in the 
value of q(TIR/Radio) throughout the SGS region (see Table~\ref{t1}). The mean 
value of ``q'' derived within the SGS region ($\simeq$ 3.0) is consistent
with, though slightly larger than, the average global values found in a sample 
of larger spiral galaxies in the \sings\ sample \citep{murphy05}; we find 
variations of $\simeq$ 1 dex in the value of ``q'' throughout the SGS region.  
These strong variations may be a result of the low dust content in IC\,2574 
(an effect of the low metal content, or of dust destruction in the extreme SGS 
environment).

%-----------------------------------------------------------------------------%
\section{Conclusions}
\label{S5}
%-----------------------------------------------------------------------------%

We have presented a multiwavelength study of the SGS region in IC\,2574, 
highlighting new \spitzer\ imaging obtained as part
of \sings. The unique multiwavelength properties demonstrate
that the expanding shell is dramatically 
affecting its surroundings by triggering SF and by altering the 
dust temperature and characteristics.

%-----------------------------------------------------------------------------%
\acknowledgements
%-----------------------------------------------------------------------------%

The {\it Spitzer Space Telescope} Legacy Science Program ``The Spitzer
Infrared Nearby Galaxies Survey'' was made possible by NASA through contract 
1224769 issued by JPL/Caltech under NASA contract 1407.  The authors thank
Evan Skillman for useful discussions, and the anonymous referee for comments 
that helped to improve the manuscript.


\begin{thebibliography}{}

\bibitem[Bell(2003)]{bell03a} Bell, E.~F.\ 2003, \apj, 586, 794

\bibitem[Bianchi \etal(1999)]{bianchi99} Bianchi, S., Davies, J.~I., \& Alton,
P.~B.\ 1999, \aap, 344, L1 

\bibitem[Dale \& Helou(2002)]{dale02} Dale, D.~A., \& Helou, G.\ 2002, \apj, 
576, 159

\bibitem[Elmegreen \& Hunter(2000)]{elmegreen00} Elmegreen, B.~G., \& Hunter, 
D.~A.\ 2000, \apj, 540, 814 

\bibitem[Gordon \etal(2005)]{gordon05} Gordon, K.D., \etal\ 2005, \pasp, 177,
503

\bibitem[Helou \etal(2004)]{helou04} Helou, G., \etal\ 2004, \apjs, 154, 253

\bibitem[Karachentsev \etal(2002)]{karachentsev02} Karachentsev, I.~D., 
\etal\ 2002, \aap, 383, 125

\bibitem[Kennicutt(1998)]{kennicutt98} Kennicutt, R.~C.\ 1998, \araa, 36, 189

\bibitem[Kennicutt \etal(2003)]{kennicutt03} Kennicutt, R.~C., \etal\ 2003, 
\pasp, 115, 928 

\bibitem[Kim \etal(1999)]{kim99} Kim, S., Dopita, M.~A., Staveley-Smith, L., 
\& Bessell, M.~S.\ 1999, \aj, 118, 2797

\bibitem[McClure-Griffiths \etal(2002)]{mccluregriffiths02} 
McClure-Griffiths, N.~M., Dickey, J.~M., Gaensler, B.~M., \& Green, A.~J.\ 
2002, \apj, 578, 176

\bibitem[Miller \& Hodge(1994)]{miller94} Miller, B.~W., \& Hodge, P.\ 1994, 
\apj, 427, 656 

\bibitem[Miller \& Hodge(1996)]{miller96} Miller, B.~W., \& Hodge, P.\ 1996, 
\apj, 458, 467

\bibitem[Murphy \etal(2005)]{murphy05} Murphy, E.~J., \etal\ 2005, 
\apj, submitted

\bibitem[S{\' a}nchez-Salcedo(2002)]{sanchezsalcedo02} S{\' a}nchez-Salcedo, 
F.~J.\ 2002, Revista Mexicana de Astronomia y Astrofisica, 38, 39 

\bibitem[Stanimirovic \etal(1999)]{stanimirovic99} Stanimirovic, S., 
Staveley-Smith, L., Dickey, J.~M., Sault, R.~J., \& Snowden, S.~L.\ 1999, 
\mnras, 302, 417

\bibitem[Stewart \& Walter(2000)]{stewart00} Stewart, S.~G., \& Walter, F.\ 
2000, \aj, 120, 1794

\bibitem[Tenorio-Tagle \& Bodenheimer(1988)]{tenoriotagle88} Tenorio-Tagle, 
G., \& Bodenheimer, P.\ 1988, \araa, 26, 145 

\bibitem[Walter(1999)]{walter99a} Walter, F.\ 1999, Publications 
of the Astronomical Society of Australia, 16, 106 

\bibitem[Walter \& Brinks(1999)]{walter99b} Walter, F., \& Brinks, E.\ 1999, 
\aj, 118, 273

\bibitem[Walter \etal(1998)]{walter98} Walter, F., Kerp, J., Duric, N., 
Brinks, E., \& Klein, U.\ 1998, \apjl, 502, L143

\bibitem[Wilcots \& Miller(1998)]{wilcots98} Wilcots, E.~M., \& Miller, 
B.~W.\ 1998, \aj, 116, 2363

\end{thebibliography}
\end{document}